\documentclass[sn-mathphys]{sn-jnl}

\jyear{2022}%

\theoremstyle{thmstyleone}%
\theoremstyle{thmstyletwo}%

\theoremstyle{thmstylethree}%

\begin{document}

\title[Use and misuse of variances for quantum systems in pure or mixed
states]{Use and misuse of variances for quantum systems in pure or mixed
states}

\author*[1]{\fnm{Alain} \sur{Deville}}\email{alain.deville@univ-amu.fr}

\author[2]{\fnm{Yannick} \sur{Deville}}\email{yannick.deville@irap.omp.eu}
\equalcont{These authors contributed equally to this work.}

\affil*[1]{\orgdiv{IM2NP UMR 7334}, \orgname{Aix-Marseille Universit\'{e}, 
CNRS}, \orgaddress{%
\city{Marseille}, \postcode{F-13397}, 
\country{France}}}

\affil[2]{\orgdiv{IRAP (Institut de
Recherche en Astrophysique et Plan\'{e}tologie)}, 
\orgname{Universit\'{e} de Toulouse, UPS, CNRS, CNES, OMP}, 
\orgaddress{%
\city{Toulouse}, \postcode{F-31400}, 
\country{France}}}

\abstract{As a consequence of the place ascribed to measurements in the postulates of quantum mechanics, if two differently prepared systems are described with
the same density operator $\rho ,$ they are said to be in the same quantum
state.\ For more than fifty years, there has been a lack of consensus about
this postulate.\ In a 2011 paper, considering variances of spin components,
Fratini and Hayrapetyan tried to show that this postulate is unjustified.
The aim of the present paper is to discuss major points in this 2011
article, and in their reply to a 2012 paper by Bodor and Diosi claiming that
their analysis was irrelevant. Facing some ambiguities or inconsistencies in
the 2011 paper and in the reply, we first try to guess their aim, then
establish results useful in this context, and finally discuss the use or
misuse of several concepts implied in this debate.}

\maketitle

\section{Introduction\label{SectionIntroduction}}

In (Weinberg 2017) the Nobel laureate Steven Weinberg explained why, in his 
\textit{Lectures on Quantum Mechanics }(Weinberg 2013, page 95), he had
written: ``\textit{My own conclusion (not universally shared) is that today
there is no interpretation of quantum mechanics that does not have serious
flaws, and that we ought to take seriously the possibility of finding some
more satisfactory other theory, to which quantum mechanics is merely a good
approximation".} J. Bricmont and S. Goldstein (Bricmont 2018) started a
discussion of (Weinberg 2017) with a list of \textit{``other very
distinguished physicists" }having ``\textit{expressed their discontent with
our present understanding (or lack thereof) of quantum mechanics". }One may,
more or less explicitly, ignore the question of the foundations of Quantum
Mechanics (QM) and use standard QM.\ However, standard QM introduces the
concept of measurement in its founding principles, and it is well known that
John Bell criticized that situation (Bell 1990)\textit{. }One may accept the
place presently given to measurements but at the same time consider that a
similar importance should be given to the concept of a preparation. And
already in (Zeh 1970) Zeh considered that ``\textit{the statistical ensemble
consisting of equal probabilities of neutrons with spin up and spin down in
the x direction cannot be distinguished by measurement from the analogous
ensemble having the spins parallel or antiparallel to the y direction. Both
ensembles, however, can be easily prepared by appropriate versions of the
Stern-Gerlach experiment. One is justified in describing both ensembles by
the same density matrix as long as the axiom of measurement is accepted.
However, \textbf{the density matrix formalism cannot be a complete
description of the ensemble, as the ensemble cannot be rederived from the
density matrix}" }(here, and in subsequent citations, no part of the\
original text was stressed). Still unaware of the 1970 paper by Zeh,
Nenashev recently claimed that the state of a system is fully described by
its density matrix (Nenashev 2016), without explaining why, if several paths
(manipulations of a physical system initially in one or another state) lead
to the same destination (described by identical results of measurements),
and if one wishes to describe the whole story from the beginning of each
journey, keeping only this final common destination should be the right
approach.

Involved in the field of Quantum Information Processing (QIP), we have developed the concepts of Blind Quantum Source Separation (BQSS) (Deville 2007, Deville 2012, Deville 2014, Deville A. 2017) and Blind Quantum Process Tomography (BQPT) (Deville 2015, Deville Y. 2017, Deville 2020, Deville A. 2017). In that context, we were led to think about the content and possible limits of the density operator (or statistical operator, or density matrix) formalism (see e.g. (Deville A. 2022, Deville Y. 2022)). In 2011, in this Journal, a model for quantum computing with initially mixed states was proposed (Siomau 2011). The same year a paper was published in \textit{Physica Scripta} (Fratini 2011), from Fratini and Hayrapetyan (denoted as FH in what follows), and entitled \textit{Underlining some limitations of the statistical formalism in quantum mechanics}. These limitations could \textit{a priori }be shown within the whole frame -postulates and concepts- of QM, or appear as a consequence of a slight justified enhancement within this frame (see e.g.
the introduction of the random-coefficient pure state concept and its
interest (Deville A. 2022, Deville Y. 2022)). From the content of (Fratini
2011), FH appear to operate within the first context, according to which
they should obey the definitions and principles of QM, and show how the
point of view of QM may present limitations. In a 2012 ArXiv article, Bodor
and Diosi (BD) (Bodor 2012) tried to show that the conclusions contained in
(Fratini 2011) were unjustified. In their 2012 ArXiv reply (Fratini 2012),
FH claimed that they had established that, in their specific instance,
involving spins 1/2, and what they called ensembles $\mathcal{A}$ and $%
\mathcal{B},$ ``\textit{the Variance obtained by analyzing one ensemble
turned out to be different from the Variance obtained by analyzing the other
ensemble. On the other hand, the density matrices of both ensembles turned
out to be the same"}, which then, according to FH, leads to a single value
of the variance in the density operator approach. That FH reply did not lead
to a final agreement between these authors. The present paper\ tries to
clarify the situation, in order to help deciding whether the conclusions
from (Fratini 2011) and (Fratini 2012) should be kept.

It is first necessary to specify two points which, in (Fratini 2011), are
implicit. FH consider spins 1/2, and, in their thought-experiment, make them
cross ``\textit{a Stern and Gerlach (SG) apparatus". }One should understand
that particles with the same \textit{magnetic moment} are manipulated, since
for instance an atom with an electron spin 1/2 magnetic moment (e.g. a
silver atom, as in the 1922 experiment by Stern and Gerlach) and an atom
with a nuclear spin 1/2 (e.g. the isotope $^{19}F$) behave quite differently
in an SG device, the magnitudes of their magnetic moments being respectively
of the order of $\mu _{B},$ the Bohr magneton, and $\mu _{N},$ the nuclear
magneton ($\mu _{B}/\mu _{N}\simeq 1836$). In the following, when we speak
of identical spins 1/2, this will\ implicitly mean\ identical magnetic
moment operators, each one proportional to a spin 1/2 operator $%
\overrightarrow{s}$. An SG device was also introduced in (Zeh 1970), without
this ambiguity because Zeh imagined a collection of \textit{neutrons.}

FH also implicitly consider that these spins are distinguishable. If one
mentally thinks of a collection of SG devices, e.g. ``\textit{with the
magnetic field along the }$\widehat{x}$\textit{\ direction}" (Fratini 2011),
the spins are distinguishable through the number given to the SG device they
cross. This is also true for \textit{``particles whose spin states are
defined along the }$\widehat{z}$\textit{\ axis" }(Fratini 2011)\textit{. }%
This is not an academic distinction: the magnetic moment of the
undistinguishable conduction electrons in a metal gives rise to the
temperature-independent Pauli paramagnetism (the spin of the electron is a
fermion), whereas in an ionic insulator, the paramagnetism of
non-interacting electron spins $1/2$ (distinguishable through the ion
carrying the spin) follows a Curie law above typically a few Kelvins.

Following FH, we will be interested in a collection of an even number $N$\
of distinguishable, non-interacting, identical (magnetic moments associated
with) spins $1/2$.\ For brevity, we will call it the spin assembly.

In (Fratini 2011), FH\ use the concept of polarization, without explicitly
defining it. The degree of polarization of a spin 1/2 can be described by
its polarization vector $\overrightarrow{a}$ (Ballentine 1998, page 173).\
The components of $\overrightarrow{a}$ are real, such that $0\leq \mid a\mid
\leq 1,$ and, in standard QM, the most general state of a spin $1/2$ is
described by the density operator $\rho =(1+$ $\overrightarrow{a}%
\overrightarrow{\sigma })/2$, where $\overrightarrow{\sigma }$ is the Pauli
vector (its components are the Pauli operators $\sigma _{x}$, $\sigma _{y}$, 
$\sigma _{z}$) and $1$ a symbolic writing for the identity operator in the
state space of the spin. The spin is in a totally unpolarized state when $%
\overrightarrow{a}=0,$ and then $\rho =1/2$. If a collection of $N$
distinguishable and independent spins is considered, total unpolarization is
realized if and only if each spin is in its totally unpolarized state. This
state is described with the density operator $I/2^{N}\ (I$: unit operator in
the state space of the spin assembly).\ 

In Section \ref{SectionOfQuantumSystemsEtc}, after a reminder\ about quantum
systems and about pure or mixed states of these systems, the so-called $%
\mathcal{A}$ and $\mathcal{B}$ ensembles introduced by FH will be presented.
The way they are defined by FH somewhat mixes the concept of a quantum
system and the one of its possible pure or mixed states, and we will have to
try and guess, from the content of (Fratini 2011), which states FH consider
when calculating a variance. Then, a result concerning mean values
calculated with the statistical operator $\rho $ will be recalled and
justified.\ It has the important consequence that the difference found by FH
in their variance calculations should \textit{a priori} be either a false or
an ill-interpreted result, which is a motivation for looking for the
origin(s) of the difference they found in their variances. In Section \ref%
{SectionUsefulResults}, mean values and variances of spin components will be
calculated in several situations, in order to help the analysis of the
results given by FH.\ The origin(s) of the differences found by FH in the
calculation of the variances will be identified in Section \ref%
{SectionOriginOfDifferenceInSH}. The question of the number of particles in
the systems introduced by FH in the calcuation of mean values, and the
presence of a reference made by FH to a book by Sakurai (Sakurai, 1994),
will be discussed in Section \ref{SectionDiscussion}.

\section{Of quantum systems, their pure or mixed sta\-tes, and mean values
in these states\label{SectionOfQuantumSystemsEtc}}

We did not try to systematically keep the notations from (Fratini 2011), as
some quantities used in our paper are absent in (Fratini 2011), and some
notations in (Fratini 2011) are ambiguous (e.g. the distinction between the
spin\ $\overrightarrow{s_{i}}$ of spin $i$ and the total spin $%
\overrightarrow{S}$ $=\sum_{i}$\ $\overrightarrow{s_{i}}$ $\ $of the spin
collection). The dimension of $\mathcal{E}$, the state space of the quantum
system of interest $\sum ,$ is supposed to be finite, denoted as $d,$ and a
basis of orthonormal kets of $\mathcal{E}$ will be denoted as the collection
\{$\mid k>$\}. The completeness relation is expressed as $\sum_{k}\mid
k><k\mid =I$ ($I$: the identity operator acting in $\mathcal{E}$). A pure
state of a spin 1/2 is usually written, in the Dirac formalism, as a ket $%
\mid \Phi >=\alpha \mid +>+\beta \mid ->$ (an element of the state space of
the spin,$\ \alpha $ and $\beta $ being complex coefficients with $\mid
\alpha \mid ^{2}+\mid \beta \mid ^{2}=1$). \ When a collection of $N$
identical distinguishable spins $1/2$ is considered, then, in order to
describe its possible pure states, the tensor product of the state space of
each spin, with dimension $d=2^{N}$, is introduced, and it is a postulate
that every ket of this space represents a pure state of the spin assembly.
A mixed state of a quantum system (e.g a single spin, or this collection of
spins), is a collection of unit-norm pure states, $\mid \Phi _{1}>$, ...$%
\mid \Phi _{i}>...$ of this system, with respective probabilities $p_{1}$%
,... $p_{i}$ ... (for each $p_{i},$\ $0\leq p_{i}\leq 1,$ and $\sum p_{i}=1$%
).\ With that mixed state one may associate the statistical operator $\rho
=\sum_{i}p_{i}\mid \Phi _{i}><\Phi _{i}\mid $, and in the specific case of a
pure state $\mid \Phi >,$ then $\rho =\mid \Phi ><\Phi \mid $, the
associated projector.

Before trying to go on clarifying the situations discussed by FH in (Fratini
2011), it is useful to make some comments about the concept of a mixed
state, accepting the description of a pure state by a ket. In his 1932 book
(1955 English translation of (von Neumann, 1932), pages 295-296), von
Neumann, having introduced the concept of a pure state (if not \ the
expression, cf. the note in the present paper), and the probability content
attached to it, presented that of a mixed state, writing \ that \textit{``the
statistical character may become even more prominent, if we do not even know
what state is actually present \ - - for example when several states }$\phi
_{1},$ \textit{\ }$\phi _{2},...$\textit{\ with the respective probabilities 
}$w_{1},$\textit{\ }$w_{2},...$($w_{1}\geq 0,$ $w_{2}\geq 0,..$.$%
w_{1}+w_{2}+...=1$) \textit{constitute the description}" of the system of
interest. The ket formalism was introduced by\ Dirac seven years later
(Dirac, 1939). Some eighty years after von Neumann's book, Weinberg used the
same ideas when writing ``\textit{Probabilities can enter in quantum
mechanics not only because of the probabilistic nature of state vectors, but
also because (just as in classical mechanics) we may not know the state of a
system. A system may be in any one of a number of states, represented by
state vectors }$\Psi _{n}$\textit{\ that are normalized }but not necessarily
orthogonal\textit{, with probabilities }$P_{n}$ \textit{satisfying }$\Sigma
_{n}P_{n}=1"$ (Weinberg 2013, page 68).$\ $In the meantime, in his 1957
review (Fano 1957), Fano had also explicitly used the concept of a pure
state when introducing mixed states, moreover writing that ``\textit{to
calculate the probability of finding a certain experimental result with a
system in the mixed state one must first calculate the probability for each
of the pure states and then take an average, attributing to each of the pure
states an assigned ``weight"".}

These passages from von Neumann and Weinberg contain an interpretation of
the mixed state concept implying an ignorance of the experimenter. A brief
comment on its possible use in the context of (Fratini 2011) will be made in
Section \ref{SectionDiscussion}.

\subsection{The ensembles $\mathcal{A}$ and $\mathcal{B}$ introduced by FH 
\label{SubSectionAandBEnsemblesFromFH}}

FH define their ensemble $\mathcal{A}$ as follows: ``$N$ \textit{particles
whose spin states are defined along the }$\widehat{x}$\textit{\ axis, }$N/2$%
\textit{\ of which are eigenstates of the spin operator }$\widehat{S}_{x}$ 
\textit{with eigenvalue }$+\hslash /2,$ \textit{while the remaining }$N/2$%
\textit{\ are eigenstates of the same operator with eigenvalue }$-\hslash
/2" $ (and FH define their ensemble $\mathcal{B}$ just by replacing \textit{%
the }$\widehat{x}$\textit{\ }axis by the $\widehat{z}$\textit{\ }axis and
their $\widehat{S}_{x}$ by their $\widehat{S}_{z}$)$.\ $This defines partly $%
\sum $ and quite partly its quantum state. $\sum $ is a collection of
particles with spin $1/2$, and the rest of (Fratini 2011) indicates that FH
consider only their spin. It is not explicitly written that they are
distinguishable, but nowhere is any antisymmetry of the fermion states
suggested, and their experimental conditions, with the use of an SG device,
clearly suggest that they can be distinguished. This passage of (Fratini
2011) is ambiguous, as ``\textit{states are defined along the }$\widehat{x}$%
\textit{\ axis" }may just refer to the choice of a basis for the description
of the spin states, and further in (Fratini 2011), FH give contradictory
precisions on the nature of the state -pure or mixed- of the spin assembly,
since they first claim that ``\textit{Both }$\mathcal{A}$\textit{\ and }$%
\mathcal{B}$\textit{\ are totally unpolarized ensembles. In the language of
statistical quantum mechanics, which will be encountered in section 4, }%
\textbf{they are also said to be in a maximally mixed state}". But it is
quite difficult to suppose that experimental conditions not detailed in
(Fratini 2011) have resulted in a totally mixed state, because precisions
then added in (Fratini 2011) rule out this supposition. More precisely, FH
write: \textit{``we can pragmatically think of using, for example, a Stern
and Gerlach (SG) apparatus with the magnetic field along the }$\widehat{x}$%
\textit{\ direction [20]. The SG apparatus would measure the spin of each
single particle of the ensemble, so that the total spin of the ensemble
would then be obtained as the sum of all the single-particle spin
measurements.",\ }and \ ``\textit{Because of the characteristics chosen for
ensemble }$\mathcal{A}$\textit{, the SG apparatus would exactly separate
particle flux into two equal parts, or, what is the same, it will measure
N/2 particles having spin along the }$+\widehat{x}$\textit{\ direction and
N/2 particles having spin along the }$-\widehat{x}$\textit{\ direction. }%
\textbf{The probability of registering the outcome }$\pm \hslash /2$\textbf{%
, when the spin along }$\widehat{x}$\textbf{\ is measured on state }$\mid 
\mathbf{S}_{x},$\textbf{\ }$\mathbf{\pm 1}>,$ \textbf{\ is in fact 1"}. The
latter citation leaves no doubt: at the input of the SG device, each spin is
either in its $\mid +x>$ or $\mid -x>$ state, and this is still true at its
output. And in their 2012 reply (Fratini 2012) to Bodor and Diosi (Bodor
2012), FH do insist that ``\textit{These two ensembles are definitely not
statistical ensembles", }adding\textit{\ ``we considered prepared ensembles". 
}If in contrast one accepts to consider that the spin assembly is in the
totally unpolarized state $\rho =I/2^{N}$, then Eq.\ (8) of (Fratini 2011)
is false (this equation states that, for what FH call their ensemble $%
\mathcal{A},$ the mean value and the variance of the component $S_{x}$ of
the total spin, which is denoted as $S_{x}^{\mathcal{A}}$ in (Fratini 2011)$%
, $ are both equal to $0,$ a result to be opposed to the true result
established below in Subsection \ref{SubsectionTotallyUnpolarizedMixture})$.$%
\ Moreover, if one tries to suppose that FH\ improperly used their
expression ``unpolarized ensemble" for their ensemble $\mathcal{A}$ in order
to describe a pure state with a mean value of their $S_{x}^{\mathcal{A}}$
equal to $0$, one has to give up this assumption when reaching Section 4 of
(Fratini 2011), in which FH consider mixed states, not pure states. In
Subsection \ref{SubsectionStillTryingToUnderstandFH}, we will suppose that
FH\ adopt the point of view of someone who \textit{ignores }that their
ensemble $\mathcal{A}$\ was prepared in a well-defined pure state, and then
decides\textit{\ to assume }that the collection of $N$ spins is in a mixed
state where the $2^{N} $ states of the $S_{x}$ eigenbasis all contribute
with the same weight to the statistical mixture $(\rho =I/2^{N}).$ We will
then show that this approach should be discarded.

Faced with these ambiguities and inconsistencies, in Section \ref%
{SectionUsefulResults} we will successively consider the spin assembly in
pure or mixed states chosen for their interest in the analysis of the
content of (Fratini 2011), without identifying specifically one of the
results with what FH intended to do.

\subsection{More on pure or mixed states\label%
{SubsectionMoreOnPureMixedStates}}

FH speak of the \textit{``total spin of ensemble }$\mathcal{A}$" (Fratini
2011). Their definition of $\mathcal{A}$ suggests us to introduce the
observable $S_{x}=\sum_{i}s_{xi}$ (we use the standard definition for the $%
\widehat{z}$ axis), a linear Hermitian operator acting in the state space $%
\mathcal{E=E}_{1}\otimes \mathcal{E}_{2}\mathcal{\otimes }...\otimes 
\mathcal{E}_{N},$ and $s_{x1}$ is a condensed writing for $s_{x1}\otimes
I_{2}..\otimes I_{N},$ where e.g.\ $I_{2}$ is the identity operator acting
in $\mathcal{E}_{2}.$ In order to make more explicit what seems to have been
FH's aim, we imagine an experiment with a spin 1/2 crossing an SG device
with the field gradient along the $\widehat{x}$ direction, its downward
(-1/2) path being interrupted by the presence of a detector. If a spin 1/2
is sent into its input and no spin is then detected at the level of the
interrupted path, in experimental conditions when the spin is not lost, one
then knows that the spin arrived at the upward aperture, and may say that
this SG device prepared this spin in the $\mid +x>$ state, where $s_{x}\mid
+x>=(1/2)\mid +x>$\thinspace\ (we systematically use reduced units, here $%
1/2 $ and not $\hslash /2$).\ We now imagine a collection of $N$ such
independent SG devices ($N$ even), each one able to prepare either in the $%
\mid +x>$ or in the $\mid -x>$ state the spin which crossed it, and $N$
spins $1/2$ simultaneously sent, one at the input of each SG device (the
spins are then distinguishable through the number of the SG device they
cross). Depending on the positions of the $N$ detectors, one may obtain $%
2^{N}$ such (pure) product-states, each one an eigenstate of $S_{x}$. One of
these $2^{N}$ product-states has all the spins in their $\mid +x>$ state,
one of them has all the spins in their $\mid -x>$\thinspace\ state, and a
number of them have $N/2$ spins in the $\mid +x>$ state (as in the states in
the ensemble $\mathcal{A}$ of (Fratini 2011)), e.g. the one with the $N/2$
first ones in the $\mid +x>$ state, and the remaining ones in the $\mid -x>\ 
$state, the following state:%
\begin{equation}
\mid 1+x,2+x...N/2-1+x,N/2+x,N/2+1-x...,N-1-x,...N-x>,\text{ \label%
{Etat1SelonOx}}
\end{equation}%
a compact writing for the tensor product of $N$ kets, each one defined by
the number of the spin, the fact that it is an eigenstate of the $x$
component of that spin, and the $+$ or $-$ state. We will call $\mid \Psi
_{x\delta }>$ the one corresponding to the specific ordered choice, called $%
\delta $, of the $N$ individual kets in a given product-state.\ When FH
speak of their ensemble $\mathcal{A}$, composed of $N$ spins (with $N$
even), and simultaneously write that the result of the measurement of the $%
\widehat{x}$ component (of a spin) is surely $+\hbar /2$ or $-\hbar /2,$
depending on the number $i$ of the measured spin, and that the numbers of $%
+\hbar /2$ and $-\hbar /2$ results are both equal to $N/2,$ this implicitly
means that they consider that the assembly is in one of the pure states just
denoted as $\mid \Psi _{x\delta }>$, and that they do not tell us what
specific ordered choice (of the $+$ and $-$ values) presently denoted as $%
\delta $ is implied.

Since FH also speak of a mixed state, we will also imagine, for the spin
assembly, the mixed state $\{\mid \Psi _{x\delta 1}>,N_{N/2}^{-1},...\mid
\Psi _{x\delta N_{N/2}^{{}}}>,N_{N/2}^{-1}\},$ each pure state implying $N/2$
spins with the $\mid +x>\ $state, and $N/2$ with $\mid -x>$, all with the
same weight $N_{N/2}^{-1}$, where $N_{N/2}=N!/((N/2)!(N/2)!)$ is the number
of distinct orderings of these $N$ spin states (cf.\ (Mathews 1965), p. 352:
if e.g. $N=4,$ $N_{N/2}$ $=6$).\ Such a mixed state therefore mobilizes the
whole collection of distinct $\mid \Psi _{x\delta }>$ pure states, all with
the same weight in the mixture.\ Of course, the statistical operator $\rho $
associated with this mixture is \textit{not }equal to $I/2^{N},$ since only
part of the eigenvectors of $S_{x}$ are of the form $\mid \Psi _{x\delta }>$
(e.g. the product-ket $\Pi _{\otimes i}\mid i,+x>$, implying the $\mid +x>$
state of each spin, is not of that form). Consequently, while it is easy to
imagine an experimental device enabling one to get the mixture described by
the statistical operator $I/2^{N}$ (think of the collection of spins in an
oven at temperature $T,$ the whole being placed in a static magnetic field,
and the Zeeman energy $E_{Z}$ verifying $E_{Z}<<kT),$ getting the mixed
state $\{\mid \Psi _{x\delta 1}>,N_{N/2}^{-1},...\mid \Psi _{x\delta
N_{N/2}^{{}}}>,N_{N/2}^{-1}\}$, necessitates some selection among the pure
states and therefore should be more elaborate.

\subsection{A consequence of the definition of the statistical operator $%
\protect\rho $ \label{SubSectionResultForRho}}

If $\Sigma $ is in a pure state described by the normed ket $\mid \Phi >,$
the mean value of an observable $\widehat{O}$ acting on the kets of $%
\mathcal{E}$\ is the quantity 
$<$%
$\Phi $ $\mid \widehat{O}$ $\mid
\Phi >.$ If $\Sigma $ is in a mixed state symbolically written \{$\mid \Phi
_{i}>,$ $p_{i}$\}, a collection of normed (but not necessarily orthogonal)
kets $\mid \Phi _{i}>,$ weighted by the respective probabilities $p_{i}$,
the mean value of $\widehat{O}$ in that mixture is the quantity:%
\begin{equation}
\sum_{i}p_{i}<\Phi _{i}\mid \widehat{O}\mid \Phi _{i}>=\sum_{i,k,k^{\prime
}}p_{i}<\Phi _{i}\mid k><k\mid \widehat{O}\mid k^{\prime }><k^{\prime }\mid
\Phi _{i}>
\end{equation}%
\ where the completeness relation was used twice. This quantity may be
written

\begin{equation}
<\widehat{O}>=\sum_{kk^{\prime }}\rho _{k^{\prime }k}O_{kk^{\prime }}\text{ %
\label{<0>=Somme}}
\end{equation}%
with%
\begin{equation}
O_{kk^{\prime }}\text{=}<k\mid \widehat{O}\mid k^{\prime }>\text{, \label%
{ElementDeMatriceDeO}}
\end{equation}%
and%
\begin{equation}
\rho _{k^{\prime }k}=\sum_{i}p_{i}<k^{\prime }\mid \Phi _{i}><\Phi _{i}\mid
k>.\text{\label{rhokkprime}}
\end{equation}%
$O_{kk^{\prime }}$ is a matrix element, in the chosen basis, of the
observable $\widehat{O}$, and $\rho _{k^{\prime }k}$ a matrix element, in
the same basis, of the statistical operator $\rho $ (see e.g. (Fano 1957)).
The mean value may then be written:%
\begin{equation}
<\widehat{O}>=\sum_{k^{\prime }}(\rho \widehat{O})_{k^{\prime }k^{\prime
}}=Tr(\rho \widehat{O})\text{ \label{<O>=Tr(rhoO)}}
\end{equation}%
The statistical matrix may be written as $\rho =\sum_{_{k,k^{\prime }}}\rho
_{kk^{\prime }}\mid k><k^{\prime }\mid ,$ and using Eq. (\ref{rhokkprime}),
one finally gets:%
\begin{equation}
\rho =\sum_{i}p_{i}\mid \Phi _{i}><\Phi _{i}\mid \text{ \label%
{ExpressionDeRho}}
\end{equation}%
Therefore, if $\Sigma $ is in a mixed state (or statistical mixture) \{$\mid
\Phi _{i}>,$ $p_{i}$\}, and if one is interested in the mean value $<%
\widehat{O}>$ or the variance $<(\widehat{O}-<\widehat{O}>)^{2}>$ of $%
\widehat{O}$, or more generally in the mean value of some function of $%
\widehat{O},$ then, \textbf{as a consequence of the definition of the
density operator }$\mathbf{\rho }\,,$\textbf{\ the result calculated using
directly the definition \{}$\mathbf{\mid \Phi }_{\mathbf{i}}\mathbf{>}_{%
\mathbf{,}}$ \textbf{\ }$\mathbf{p}_{\mathbf{i}}$\textbf{\} of the mixed
state, and the result obtained using the statistical operator }$\mathbf{\rho 
}$\textbf{\ and the trace are necessarily equal. \ }This well-known result,
ignored in (Fratini, 2011), will be explicitly used in Section \ref%
{SectionDiscussion}.

For the definition of a variance in a quantum context, one can see e.g.
(Ballentine 1998) or (Ohya 2011). Robertson, when introducing the variance
in a quantum context in (Robertson 1929), called the square root of the
variance \textit{the uncertainty.}

\section{Spin components of the spin assembly: useful results\label%
{SectionUsefulResults}}

We now calculate mean values and variances of spin components of the spin
assembly, in some specific pure or mixed states. They will be used in the
next section, when trying to explain the origin of the differences found in
the values of variances in the 2011 paper by FH.

\subsection{Spin assembly in the pure state $\mid \Psi _{x\protect\delta }>$%
\label{SubsectionSpinAssemblyPureStatePisxdelta}}

\subsubsection{The $S_{x}$ component, and pure state $\mid \Psi _{x\protect%
\delta }>$\label{SubsectionSxComponentAndPureStatePsixDelta}}

When the spin assembly is in the pure state $\mid \Psi _{x\delta }>$,
calculating the mean value and the variance of $S_{x}$ is easy because:\ 1) $%
S_{x}$ is the sum of all $s_{xi}~$components, $\ $2) $\mid \Psi _{x\delta }>$
is a tensor product without partial entanglement, 3) each factor of the
product is an eigenket of the corresponding $s_{xi}$, and the eigenvalues
may be written $\varepsilon _{i}/2,$ with $\varepsilon _{i}=\pm 1.$\
Consequently, for any ordering defined by $\delta $:%
\begin{equation}
S_{x}\mid \Psi _{x\delta }>=\frac{1}{2}\sum_{j=1}^{N}\varepsilon _{j}\mid
\Psi _{x\delta }>=0
\end{equation}%
because, in the sum $\sum_{j=1}^{N}\varepsilon _{j},~$with $\varepsilon
_{j}=\pm 1,$ the values $1$ and $-1$ both appear $N/2$ times. The mean value
of the total spin $S_{x}$ in any of these pure states is:%
\begin{equation}
<S_{x}>=<\Psi _{x\delta }\mid S_{x}\mid \Psi _{x\delta }>=0
\end{equation}%
since $S_{x}\mid \Psi _{\emph{x}\delta }>$ is equal to zero. The variance of
the total spin $S_{x}$ in any of these pure states is: 
\begin{equation}
<\Psi _{x\delta }\mid (S_{x}-<S_{x}>)^{2}\mid \Psi _{x\delta }>=<\Psi
_{x\delta }\mid S_{x}^{2}\mid \Psi _{x\delta }>=0
\end{equation}%
which results from the fact that presently: 1) $<S_{x}>=0,$ 2) $%
S_{x}^{2}\mid \Psi _{x\delta }>=$ $S_{x}^{{}}(S_{x}\mid \Psi _{x\delta }>),$
which equals $0,$ as $S_{x}\mid \Psi _{x\delta }>=0.$

Both the mean value and the variance of $S_{x}$ are therefore equal to zero.
To sum up, $S_{x}$ has a zero variance because $\mid \Psi _{x\delta }>$ is
an eigenvector of $S_{x}$ for the eigenvalue $0.$

Calculating the mean value and the variance of $S_{x}$ was easy, but it
would be wrong to think that the assembly could be replaced by a single
spin. $\mid \Psi _{x\delta }>$ describes a pure state of an $N$ spin $1/2$
collection, and e.g. the just used sums $\sum_{j=1}^{N}s_{jx}$ and $%
\sum_{j=1}^{N}\varepsilon _{j}\,$\ mobilize the whole assembly. For
instance, if $\mid \Psi _{x\delta }>$ is a state with the first spin in the $%
\mid +x>$ state, the mean value $<1+x\mid s_{1x}\mid 1+x>$ is equal to $%
+1/2, $ and this is true, in this specific case, for $<\Psi _{x\delta }\mid
s_{1x}\mid \Psi _{x\delta }>$. But $<\Psi _{x\delta }\mid S_{x}\mid \Psi
_{x\delta }>,$ the mean value of the $x$ component of the total spin, is a
sum of $N$ such quantities, $N/2$ of them being equal to $+1/2$ and the $N/2$
remaining ones being equal to $-1/2.$

\subsubsection{The $S_{z}$ component, and\ again pure state $\mid \Psi _{x%
\protect\delta }>$}

We first consider\ a single spin $1/2$, in the pure state $\mid \pm x>:$
when acting on $\mid \pm x>,$ the basis vectors of $s_{x}$, $s_{z}$
generates the vectors $(1/2)\mid \mp x>$, and its mean value in $\mid \pm x>$
is\hspace{-2cm}%
\begin{eqnarray}
&<&\pm x\mid s_{z}\mid \pm x>=<\pm x\mid s_{z}\frac{\mid +>\pm \mid ->}{%
\sqrt{2}} \\
&=&<\pm x\mid \frac{1}{2}\frac{\mid +>\mp \mid ->}{\sqrt{2}}=\frac{1}{2}<\pm
x\mid \mp x>=0
\end{eqnarray}%
Its variance in this state is presently equal to the mean value of $%
s_{z}^{2},$ i.e. to%
\begin{equation}
<\pm x\mid s_{z}^{2}\mid \pm x>=\frac{1}{4}
\end{equation}%
as $s_{z}^{2}\mid \pm x>$ $=s_{z}(s_{z}\mid \pm x>)=s_{z}(1/2)\mid \mp
x>=(1/4)\mid \pm x>.$

We now come to the mean value and variance of $S_{z}$ in the pure state $%
\mid \Psi _{x\delta }>.$\ The mean value of $S_{z}=\sum_{j}s_{jz}$ is a sum
of $N$ contributions, each one equal to $0$, and therefore $<S_{z}>=<\Psi
_{x\delta }\mid S_{z}\mid \Psi _{x\delta }>=0.$

The variance of the total spin component $S_{z}$ is: 
\begin{equation}
<\Psi _{x\delta }\mid (S_{z}-<S_{z}>)^{2}\mid \Psi _{x\delta }>=<\Psi
_{x\delta }\mid S_{z}^{2}\mid \Psi _{x\delta }>
\end{equation}%
$S_{z}^{2}$ introduces two sorts of terms: the $s_{iz}s_{jz(j\neq i)}$
terms, which do not contribute, and the $N$ terms of the form $s_{iz}^{2}$,
each making a contribution equal to $1/4$:%
\begin{equation*}
<\Psi _{x\delta }\mid (S_{z}-<S_{z}>)^{2}\mid \Psi _{x\delta }>=N/4
\end{equation*}

\subsection{S$_{x}$ and the mixed state $\{\mid \Psi _{x\protect\delta %
1}>,N_{N/2}^{-1},...\mid \Psi _{x\protect\delta N_{N/2}^{{}}}>,N_{N/2}^{-1}%
\} $\label{SubSectionMixedStateCalculations}}

The spin assembly is now supposed to be in the mixed state $\{\mid \Psi
_{x\delta 1}>,N_{N/2}^{-1},...$ $\mid \Psi _{x\delta
N_{N/2}^{{}}}>,N_{N/2}^{-1}\},$ the way this mixed state was obtained not
being described. As a first step in the calculation of the mean value and
the variance of the total spin $S_{x}$, one has to consider the contribution
of a single pure state $\mid \Psi _{x\delta }>$ of the mixture \ to this
mean value and to this variance.\ Here, the result is simple: from the
result obtained in Subsection \ref{SubsectionSpinAssemblyPureStatePisxdelta}
both contributions are equal to zero for any $\delta $ ordering.\ Therefore,
the mean value and the variance of $S_{x}$ in this mixed state, the sum of
the corresponding $N_{N/2}^{{}}$ contributions, are again both equal to $0.$

\subsection{The totally unpolarized mixture\label%
{SubsectionTotallyUnpolarizedMixture}}

If the spin assembly is in the totally unpolarized mixed state $\rho
=I/2^{N},$ and if one wishes to calculate the mean value and the variance of 
$S_{x},$ this mixed state may be interpreted as made up of the $2^{N}$
tensor products $\mid 1\pm x>\otimes \mid 2\pm x>\otimes ...\otimes \mid
N\pm x>$, a basis of eigenstates of $S_{x},$ all with the same weight $%
1/2^{N}$. The mean value of $S_{x}$ is equal to $0$ since: 1) $S_{x}$ is a
sum of $N$ terms, each one implying a single spin, 2) the contribution of
any such term to the Trace of $S_{x}$, calculated in the eigenbasis of $%
S_{x}\,$,$\,$is $0,\ $as there are $N/2$ diagonal elements equal to $1/2$
and $N/2$ diagonal elements equal to $-1/2$. The variance of $S_{x}$ is here
equal to $N/4,$ as the $s_{xi}s_{xj(j\neq i)}$ terms do not contribute, and
each of the $N$ terms of the form $s_{xi}^{2}$ in the sum $%
\sum_{i}s_{xi}^{2} $ makes a contribution to the trace of $\rho S_{x}^{2}$
equal to $(1/2^{N})2^{N}(1/4)$).\ \ The same results are obtained for $S_{z}$%
, using the standard basis, and the kets\ $\mid 1\pm >\otimes \mid 2\pm
>\otimes ...\otimes \mid N\pm >:$%
\begin{equation*}
\text{If }\rho =\frac{I}{2^{N}}\text{ (totally unpolarized mixture): }%
\left\{ 
\begin{array}{c}
\text{mean value of }S_{x}=0 \\ 
\text{variance of }S_{x}=\frac{N}{4}%
\end{array}%
\right.
\end{equation*}%
and the same is true for $S_{z}.$

\subsection{Exchanging the roles of $x$ and $z$\label%
{SubsectionExchangingxand z}}

The existence and the use of the ensemble $\mathcal{B}$\ in\ (Fratini 2011)
suggest us to consider quantum states, either pure or mixed, and
observables, obtained from each situation already considered in this
section, and to calculate the corresponding mean values and variances, once
the $\widehat{x}$ axis has been replaced by the $\widehat{z}$ axis, and the
total spin component $S_{z}$ and the pure state $\mid \Psi _{z\delta }>$
have replaced $S_{x}$ and $\mid \Psi _{x\delta }>.\ $Considering e.g. $\mid
\Psi _{z\delta }>,$ for a similar reason to the one found with $S_{x}$ and $%
\mid \Psi _{x\delta }>$, now $S_{z}\mid \Psi _{z\delta }>=0$, and the mean
value and the variance of $S_{z}$ in a $\mid \Psi _{z\delta }>$ state are
equal to zero. Had a difference been found for the results for the $\widehat{%
x}$ axis with the total spin component $S_{x}$ on one side, and the $%
\widehat{z}$ axis with the total spin component $S_{z}$ on the other, this
would have meant that the isotropy of space was not respected.

\section{The origins of the difference in the values of variances found by
FH \label{SectionOriginOfDifferenceInSH}}

In Section \ref{SectionUsefulResults} we found that the mean value and the
variance of $S_{x}$ are both equal to zero, in both the pure state $\mid
\Psi _{x\delta }>,$ for an arbitrary $\delta $ ordering, and the mixed state 
$\{\mid \Psi _{x\delta 1}>,N_{N/2}^{-1},...\mid \Psi _{x\delta
N_{N/2}^{{}}}>,N_{N/2}^{-1}\}.$\ The origin(s) of the difference in the
values of variances claimed by FH must be identified and explained in some
detail. This is not an easy task, as we showed that their (Fratini 2011)
paper contains ambiguous and even contradictory elements.\ \textbf{We first
consider the results they get in their Section 3}. FH calculate the variance
of the same observable, first for their ensemble $\mathcal{A}$ (their Eq.
(8)) and then for their ensemble $\mathcal{B}$ (their Eq. (13)), find that
the values differ and then write ``\textit{We can certainly conclude that the
two ensembles }$\mathcal{A}$\textit{\ and }$\mathcal{B}$\textit{\ are not
equal".\ }\ Speaking of \textit{ensembles which are} \textit{not equal }is
using an expression which is not defined within QM. In fact, two quantum
systems may be identical or distinct, and if they are identical, they may be
in the same state or not. In Section \ref{SectionIntroduction} we gave
reasons why the system considered by FH in (Fratini 2011) is a collection of
distinguishable, independent, identical magnetic moments associated with
spins $1/2$. In Subsection \ref{SubSectionAandBEnsemblesFromFH} we stressed
that (Fratini 2011) and (Fratini 2012) by FH contain ambiguities and
inconsistencies. We then introduced a pure state of the spin assembly which
we called $\mid \Psi _{x\delta }>$ and gave reasons why, when FH speak of 
\textit{ensemble }$\mathcal{A}$, it should be understood that the system is
in this pure state $\mid \Psi _{x\delta }>$.\ In Subsection 3.1 of (Fratini
2011), FH speak of ``\textit{the total spin of ensemble }$\mathcal{A}$\textit{%
"},$\ $and the existence of their Eq. (11), with the presence of $N,$ the
number of spins, clearly confirms that, whereas their notation is unclear,
they are interested in the mean value and the variance of the $\mathbf{x}$%
\textit{\ component} of the \textit{total }spin. It is then interesting to
examine the following two situations: \textbf{one first considers their
ensemble }$\mathcal{A}$ \textbf{only,} and the mean value and the variance
of $S_{z}=\sum_{i}s_{zi}$\textbf{\ in the pure state }$\mid \mathbf{\Psi }_{%
\mathbf{x\delta }}>:$ one gets (cf.\ Section \ref{SectionUsefulResults})%
\begin{equation}
<S_{z}>=0,\text{ \ \ \ \ \ \ }<(S_{z}-<S_{z}>)^{2}>=\frac{N}{4},
\end{equation}%
to be opposed to%
\begin{equation}
<S_{x}>=0,\text{ \ \ \ \ \ \ }<(S_{x}-<S_{x}>)^{2}>=0\text{\ \ \ \ }
\end{equation}%
\textbf{One then considers their ensemble }$\mathcal{B}$ \textbf{only:} the
roles of $S_{x}$ and $S_{z}$ have only to be exchanged, as explained in
Subsection \ref{SubsectionExchangingxand z}, and if one defines $\mid \Psi
_{z\delta }>$ by replacing the kets $\mid \pm x>$ by the kets $\mid \pm >$
(here keeping the usual notation), the mean values \textbf{in the pure state 
}$\mid \mathbf{\Psi }_{\mathbf{z\delta }}>$ are respectively:%
\begin{equation}
<S_{x}>=0,\text{ \ \ \ \ \ \ }<(S_{x}-<S_{x}>)^{2}>=\frac{N}{4}
\end{equation}%
\begin{equation}
<S_{z}>=0,\text{ \ \ \ \ \ \ }<(S_{z}-<S_{z}>)^{2}>=0
\end{equation}%
Therefore the mean value and the variance of $S_{\mathbf{x}},$ instead of
those of $S_{z},$ are $0$ and $N/4$ respectively, as can also be obtained
using a projector and a Trace (cf. Subsection \ref{SubSectionResultForRho}).
The difference has two origins: 1) $S_{x}$ and $S_{z}$ do not commute, and
play symmetrical roles, $2)~$one compares results in the pure state $\mid
\Psi _{x\delta }>$ and in the pure state $\mid \Psi _{z\delta }>.$

We then conclude, or rather suggest that: 1) $\mathcal{A}$ and $\mathcal{B}$
correspond to two \textit{distinct pure states} of the collection of $N$
(distinguishable, identical) spins, which we respectively denoted as $\mid
\Psi _{x\delta }>$ and $\mid \Psi _{z\delta }>,$ 2) the difference found by
FH in their Section 3 is due to the fact that $\mid \Psi _{x\delta }>$ is an
eigenstate of$~S_{x}$ (see\ what they call their ensemble $\mathcal{A}$),
but not of $S_{z}$, whereas $\mid \Psi _{z\delta }>$ is an eigenstate of $%
S_{z}$ (see\ what they call their ensemble $\mathcal{B}$), but not of $S_{x}$%
.\ And the value of the variance of $S_{z}$ in the first case and of $S_{x}$
in the second one, $N/4$, comes from the fact that $S_{z}=\sum_{i}s_{zi}$
and $S_{x}=\sum_{i}s_{xi}$ are two sums implying $N$ spin components, each
related to a single spin.\ This value $N/4\,$is an exact result, not an
approximate one. If the spin assembly is in the state $\mid \Psi _{x\delta
}> $ and if one is interested in the mean of $S_{z}$ then, focusing on the
contribution of $s_{zi},$ one knows that in the tensor product, such as (\ref%
{Etat1SelonOx}), one may e.g. write%
\begin{equation}
\mid i,\pm x>=\frac{\mid i,+>\pm \mid i,->}{\sqrt{2}}
\end{equation}%
and if one measures $s_{zi},$ then the \textit{probability} of getting $+1/2$
is $1/2$, and that of getting $-1/2$ is also $1/2$.\ This expresses the
principle of superposition in the present situation.

This being said, \textbf{one now has to examine the content and results of
Section 4 of (Fratini 2011)}.\ FH denote the density operator $\rho $ as $%
\widehat{P}.$ If they want to be consistent with what they did in their
Section 3, they should consider, for their ensemble $\mathcal{A}$ with
(even) $N$\textit{\ spins}, the \textit{pure state} $\mid \Psi _{x\delta }>,$
and the result obtained with $\mid \Psi _{x\delta }><\Psi _{x\delta }\mid $
will be equal to the one obtained when using the ket $\mid \Psi _{x\delta }>$
(cf. Subsection \ref{SubSectionResultForRho}). Through their Eq. (16), they
instead consider a \textit{single} spin in the\textit{\ mixed} state
written, with our notations, as%
\begin{equation*}
\{\frac{1}{2}\text{,}\mid +x>,\text{ }\frac{1}{2},\mid -x>\}
\end{equation*}%
Clearly, the mean value or the variance of $s_{x}$ in this\textit{\ mixed
state of a single spin} has strictly no reason why it should be equal
respectively to the mean value or the variance of the\textit{\ total spin
component} $S_{x}=\sum s_{xi}$ in the \textit{pure state} $\mid \Psi
_{x\delta }>$ of the collection of $N$\textit{\ spins}. Their approach is
therefore inconsistent, and when, after their Eq. (19), they write that ``%
\textit{These results are in stark contrast with equations (8) and (13)", }%
their comparison is meaningless.

We are therefore justified in speaking of the origins, rather than the
origin, of the difference in the values of variances found by FH, as there
is a double inconsistency in their comparison: 1) a pure state versus a
mixed state, 2) a spin assembly ($N$ spins, with $N$ even) and a single spin.

\section{Discussion \label{SectionDiscussion}}

In the previous sections, we started from the content of the 2011 FH paper
(Fratini 2011), and tried to guess the aim of its authors.\ Some points
which were then deferred will be treated in the following discussion.

\subsection{Calculations of variances in a pure state\label%
{SubsectionCalculationVarianceInPureState}}

Whereas the deep content of QM is still presently largely discussed, through
the expression \textit{interpretations of QM (}cf. Section \ref%
{SectionIntroduction}), there is a consensus on the general rules of QM. We
examine the following simple situation: a spin 1/2 in the pure state 
\begin{equation*}
\mid V>=\frac{\mid +>+i\sqrt{3}\mid ->}{2}
\end{equation*}%
If this spin is in a pure state and if one then measures $s_{y}$,~the result
is, in a random way, either $+1/2$ or -$1/2.\ $ If the experiment consisting
of a preparation in state $\mid V>$, followed by the measurement of $s_{y}$,
is repeated $M$ times with $M>>1$, one can get (a reasonable estimation of)
the probability of occurrence of the result $+1/2$ or\textit{\ }$-1/2$.\
Using these probabilities in order to \textit{calculate} the mean value $%
<V\mid s_{y}\mid V>$ would implicitly mobilize the \{$\mid +y>$, $\mid -y>$%
\} basis. In fact, calculations are often made \textit{in a single chosen
basis, }and if, e.g.\ one uses the standard basis $\mid +>$, $\mid ->,$ and
one considers $s_{y}$ (or $s_{x}$), an observable which does not commute
with $s_{z}$, then, if one has to calculate the mean value of $s_{y}$ in
state $\mid V>$, it happens that the so-called interference terms $<+\mid
s_{y}\mid ->$ and $<-\mid s_{y}\mid +>$ contribute to this mean value, a
manifestation of the principle of superposition. One finally gets $<V\mid
s_{y}\mid V>=\sqrt{3}/4$.

The variance of $s_{y}$ in this state is (cf.\ Subsection \ref%
{SubSectionResultForRho}) the following mean value (in this state):%
\begin{equation*}
<V\mid (s_{y}-<s_{y}>)^{2}\mid V>=<V\mid s_{y}^{2}\mid V>-(<V\mid s_{y}\mid
V>)^{2}=\frac{1}{16}
\end{equation*}%
If, keeping state $\mid V>$ and the same basis, one calculates the mean
value and the variance of $s_{z}$, diagonal in the \{$\mid +>,$ $\mid ->$\}
basis, instead of those of $s_{y}$, no interference term \ appears.\ If Eq.
(4) from FH in (Fratini 2011) is valid for a first observable $\widehat{O}$,
then it cannot be used, keeping the same basis, for a second observable
which does not commute with $\widehat{O}$.\ For instance, given a pure state
of a spin $1/2,$ e.g. $\mid V>$, it cannot be used for both $s_{x}$ and $%
s_{z}.$

If one must calculate the mean value or the variance of an observable
attached to a system in a mixed state, each pure state of the mixture brings
its own contribution, and the principle of superposition operates for each
such pure state. The mean value or the variance may be calculated either by
summing the contributions of all states in the mixture (and the calculation
of each contribution mobilizes the weight affecting this pure state), or by
using the density operator, a tool for this kind of calculation, as a result
of its building up, as shown in Subsection \ref{SubSectionResultForRho}.\
Then, if in a given use the two methods do not give the same results, at
least one mistake has been made.

In (Fratini 2011), speaking of their ensembles $\mathcal{A}$ and $\mathcal{B}
$, FH write: ``\textit{We shall focus on deriving, for both ensembles, the
expectation value and the variance of the spin along the }$\widehat{x}$ 
\textit{direction. Both of them are measurable quantities}". In fact, QM
introduces so-called measurable quantities (observables), and associates
Hermitian operators with them. $s_{x}$ (or $s_{z}$) is (associated with) an
observable.\ Given a pure or mixed state, the expectation value or the
variance of $s_{x}$ (or $s_{z}$) can be calculated (with the already
recalled precautions), but they are not \textit{measurable quantities.}

In their reply (Fratini 2012) to (Bodor 2012), FH insist that their\ ``%
\textit{two ensembles are definitely not statistical ensembles}", and they
add ``\textit{we considered prepared ensembles, because we are basically free
to decide which case study to investigate".\ }In that paper by FH, the
expression ``\textit{pure state" }does not appear even once, but QM currently
considers that a pure state results from some preparation act, and one
should therefore understand that FH here claim that they consider a spin
system in a pure state. It is now possible to comment the fact that, in the
same paper (Fratini 2012), in a note in page 3, FH add that ``\textit{In Ref.
[1], the Variance has been computed by taking the quantum mechanical
prediction on the single-particles measurements and by then applying
Classical Statistics}".\ From what has been already detailed in our present
paper: 1) applying classical statistics is meaningful only if one has a
mixed state, and if quantum calculations have first been performed at the
level of each pure state.\ In (Fratini 2012), FH eliminate the case of mixed
states, 2) in the pure states FH consider in (Fratini 2011), some spin
operators have a variance containing a factor equal to $N,$ which means that
their system contains $N$ spins, and that the spin operator they consider is
a component not of a single spin, but of the total spin, which we denoted as 
$S_{z}=\sum_{i}s_{zi}$ or $S_{x}=\sum_{i}s_{xi}.$\ This being said, \textit{%
it can't be claimed} that in the calculation of the variance one should
apply \textit{Classical Statistics.\ }It is true that, given e.g. the pure
states which we called $\mid \Psi _{x\delta }>$ and $\mid \Psi _{z\delta }>,$
the calculation of the mean value of e.g. $S_{z}$ mobilizes a sum of $N$
simpler quantities. But the calculation of each of them obeys the usual
quantum rules, rather than mobilizing classical probabilities.

\subsection{FH and the Avogadro number}

We now consider the issue of the reference to the Avogadro number in the
second paper by FH (Fratini 2012).\ $N$ is the number of (identical)
particles in the physical system.\ The fact that the Avogadro number $N_{A}$
is such that $\sqrt{N_{A}}>>1$ has well-known experimental consequences, and
first in a non-quantum context, e.g. in the fact that fluctuations are very
small, except near phase transitions. And phenomenological thermodynamics
considers macroscopic systems. Our theoretical results in Section \ref%
{SectionUsefulResults} are obtained through considerations implicitly making
use of probabilities, not estimations, and they are true even for $N=2$ or $%
4.$\ In their 2012 reply FH write that ``\textit{In realistic ensembles, N is
of the order of the Avogadro Number }($\simeq 10^{23}$)" (Fratini 2012). Two
practical instances show that this is not true: in the case of electron
spins and ESR experiments, already fifty years ago, under what was called
standard conditions (which implied diluted samples in order to have weak
dipolar broadening, and non-saturating conditions for the ESR signal) it was
possible both to detect roughly $10^{13}$ spins at $300$ $K,$ and therefore
even a weaker number of them at $4.2$ $K,$ and to describe them
theoretically with QM. And roughly twenty years ago single spin ESR through
scanning tunneling spectroscopy was developed (Buchachenko 2001). In the
just mentioned passage from (Fratini 2012), FH add that (\textit{in
realistic ensembles) ``neither writing nor dealing with its N-particle
density matrix is clearly feasible".} But, precisely, in the present paper,
we did such calculations, for both the mean value and the variance of spin
components, which were possible because the spins were independent, and the
results were exact.\ Two mixed states were considered: 1) the totally
unpolarized case ($\rho =I/2^{N})$, 2) the mixed state $\{\mid \Psi
_{x\delta 1}>,N_{N/2}^{-1},...\mid \Psi _{x\delta
N_{N/2}^{{}}}>,N_{N/2}^{-1}\}$ (or, of course, the one obtained by writing $z
$ instead of $x$). The density operator $\rho $ operates in the $2^{N}$%
-dimensional state space of the spin assembly, and should not be confused
with a spin operator operating in the state space of a given spin, even in
the specific case of total unpolarization: $\rho =I/2^{N}$ is then a
(tensor) product of $N$ identity operators acting in the state space of each
spin. And, in that case of independent spins, calculations are possible
without writing explicitly the matrix associated with $\rho =I/2^{N}$ in a
basis in which $\rho $ is diagonal.

\subsection{Still trying to understand the results from FH\label%
{SubsectionStillTryingToUnderstandFH}}

We now come back to a question which was approached at the beginning of
Section \ref{SectionOfQuantumSystemsEtc}. In (Fratini 2011), FH describe a
process clearly leading to a pure state, but, as mentioned in Subsection \ref%
{SubSectionAandBEnsemblesFromFH}, they give contradictory indications on
this question, as they also suggest that their $\mathcal{A}$ and $\mathcal{B}
$ are in a mixed state (cf.\ their writing ``\textit{Both} $\mathcal{A}$ 
\textit{and} $\mathcal{B}$" ... ``\textit{are also said to be in a maximally
mixed state}"). We may suppose that they adopt the point of view of someone
who \textit{ignores }that their ensemble $\mathcal{A}$\ was prepared in a
well-defined pure state, and then \textit{supposes} that the collection of $N
$ spins is in a mixed state where the $2^{N}$ states of the $S_{x}$
eigenbasis all contribute with the same weight to the statistical mixture $%
(\rho =I/2^{N}).$ This supposition should however be rejected, since as
mentioned in Subsection \ref{SubSectionAandBEnsemblesFromFH}\ it is
inconsistent with Eq.\ (8) of (Fratini 2011).

One may instead imagine a scientist who, again ignoring that ensemble $%
\mathcal{A}$\ was prepared in an eigenstate of $S_{x}$ with the eigenvalue $%
0,$ $\mid \Psi _{x\delta }>$, decides that it is in a mixed state where the
eigenstates of the $S_{x}$ eigenbasis with this eigenvalue equal to $0$ all
contribute, with the same weight. Then, if he calculates the mean value of $%
S_{x}$ and its variance (cf.\ Subsection \ref%
{SubSectionMixedStateCalculations}), he will \textit{accidentally} find the
same results as those obtained for the pure state $\mid \Psi _{x\delta }>$
(cf. Subsubsection \ref{SubsectionSxComponentAndPureStatePsixDelta}). If he
starts from ensemble $\mathcal{B}$\emph{\ }and\emph{\ }$\mid \Psi _{z\delta
}>,$ and similarly introduces the corresponding mixed state, the same can be
said using $S_{z}.$ One could assume that this reasoning was perhaps held by
FH.\ If so, they made a confusion between the pure state $\mid \Psi
_{x\delta }>$ and a mixed state exclusively made up of states with the
eigenvalue $0$ of $S_{x},$ all with the same weight. But\textbf{\ this
explanation cannot be kept}, because, as already mentioned in Subsection \ %
\ref{SubSectionAandBEnsemblesFromFH}, in their 2012 reply (Fratini 2012),
FH\ insisted that, concerning their $\mathcal{A}$ and $\mathcal{B}$ $\ $%
ensembles: ``\textit{These two ensembles are definitely not statistical
ensembles", }adding \textit{``we considered prepared ensembles}", which may
be read as meaning that they considered pure states.

\subsection{FH and a book from Sakurai}

In both (Fratini 2011) and (Fratini 2012) FH refer to a book on quantum
mechanics by Sakurai (Sakurai 1994). This fact necessitates a short
incursion into a question belonging to the area of the interpretation(s) of
quantum mechanics. It seems that, as e.g. Ballentine (Ballentine 1998),
Sakurai adopted what Ballentine calls the ensemble interpretation (for more
details, see e.g. (Ballentine 1998)).\ But, perhaps for personal pedagogical
reasons, and contrary to Ballentine, in his book Sakurai did not use the
expression \textit{``pure state"}. Instead, he spoke of a ``\textit{state ket}%
", and introduced a ``\textit{pure ensemble"}, which he defined as \textit{``-
a collection - of identically prepared physical systems, all characterized
by the same ket }$\mid \alpha >$\textit{"}. This led him to introduce the
expression of the mean value of an observable $A$ ``\textit{taken with
respect to state }$\mid \alpha >"$ as the sum of measured values $a^{\prime
\prime }$, each one multiplied by a quantity $\mid <a^{^{\prime }}\mid
\alpha >\mid ^{2}$, interpreted as a probability (cf. his Eq. (1.4.6), page
25), a result obtained under the condition that the kets $\mid a^{\prime }>$
are (a basis of) eigenkets of $A.$\ \textbf{Therefore, if one considers two
non-commuting observables, care must be taken, as this expression cannot be
used for both of them with the same basis}. The approach followed by Sakurai
has the effect, both an advantage and a risk, of avoiding the presence of
interference terms. When reading Sakurai, one must also notice that if he
was interested in a single spin $1/2,$ in a pure state $\mid \alpha >$, he
introduced the ``\textit{number of dimensions, N, of the ket space", }equal
to $2$ for a spin\textit{\ }$1/2$, and not to be confused with the number of
spins in the ensemble associated with this spin by Sakurai. In this context
introduced by Sakurai, it is meaningless to speak of the total spin of this
collection. Sakurai described mixed states in his Subsection 3.4. (page
178). Then, what is the choice made by FH? They cite Sakurai in their two
papers, they do not speak of a system in a pure state, they speak of an
ensemble - their $\mathcal{A}$ or $\mathcal{B}$ ensemble -, but however not
of a \textit{pure ensemble} as Sakurai did, and they do not follow Sakurai's
definition, since all the members of their ensemble have not been
identically prepared, and moreover they speak of the \textit{total spin }of
their ensemble, which is contradictory with what Sakurai did. The question
is then: do FH, who cite Sakurai in their two papers, adopt his approach? If
the answer is yes, they are speaking of a single spin 1/2, and when they
speak of a total spin, this is meaningless. In spite of their reference to
Sakurai, the answer rather seems to be no, and as already said, the way they
speak of their \textit{ensemble }$A$\textit{\ }or $\mathcal{B}$\ is\textit{\ 
}ambiguous, and in the previous sections we tried to give a meaning both to
the system and its state, as possibly implicitly used by FH.

\subsection{Before concluding}

Arrived here, it should be clear that, in the work proposed by FH, some
major points are wrong and others are ambiguous and then difficult to
analyze. For instance a given result they obtain for the value of a variance
may be exact by accident, or may be wrong either because the system and its
state are ill-defined, and/or because the method they suggest (they use?) in
the calculation of an expectation value and/or a variance is wrong. And
there is a strong ambiguity upon what FH call $\mathcal{A}$ or $\mathcal{B}$
ensemble, as in their reply (Fratini 2012) they seem to insist upon the fact
they consider what is usually called a pure state, but just in contrast, in
their previous paper (Fratini 2011), they wrote that they were unpolarized 
(cf. its Section 2: ``\textit{Both} $\mathcal{A}$ \textit{and} $\mathcal{B}$ 
\textit{are totally unpolarized ensembles" }and its Section 5: ``\textit{the
two ensembles considered in this paper are both unpolarized}"), therefore 
\textit{not} in a pure state.

Faced with this reality, and\ with the following passage from BD in (Bodor
2012): ``\textit{The mistake of FH has nothing to do with quantum mechanics
rather it is a classical statistical misconception", }we consider that it
would be quite hazardous to try and give a meaning to this passage from the
reply (Fratini 2012) by FH: ``\textit{By \textquotedblleft statistical
ensemble\textquotedblright\ is meant an ensemble whose populations of states
are statistically determined by means of a certain statistical distribution
which guarantees random mixing", }\ without any reference to pure states.

\section{Conclusion}

In his 1970 paper, Zeh stressed the existence of a possible weakness in the
general postulates of QM, when observing that the density matrix formalism
cannot be a complete description of a so-called quantum ensemble, as it may
happen that this ensemble cannot be rederived from the density matrix. Some
forty years later, the 2011 paper by Fratini and Hayrapetyan (FH) aimed at
discussing possible limits of the statistical formalism in QM.\ But an idea\
is unfortunately weakened if advocated through the use of false arguments.
In the present paper, it has been shown that the differences in the values
of variances as claimed by FH are wrong results, partly as a consequence of
ill-defined and/or contradictory situations, in their definition of both a
quantum system and its possible quantum states, partly as an undue
comparison between results obtained in different situations and systems (one
spin in a mixed state versus a spin assembly in a pure state). One interest
of the paper by Bodor and Diosi is the existence of reply (Fratini 2012).\
In that paper, FH explicitly write ``\textit{The analysis we carried out in
[1] does not include nor mention statistical ensembles, as we explicitly
define and refer only to \textquotedblleft prepared
ensembles\textquotedblright }"\textit{. }If FH mean that they just
manipulated pure states, this means that when using the density matrix
formalism they have just to use projectors, and any variance calculated
through a projector has the same value as if calculated directly.\ In their
2012 reply, FH claim that ``\textit{density matrices do not provide a
complete description of ensembles of states in quantum mechanics".\ }This is
perhaps true, but FH\ failed showing it, and therefore the 2011-2012
discussion initiated by Fratini and Hayrapetyan does not affect the 1970
comment by Zeh, which therefore integrally keeps its own interest.

\section*{Declarations}

\begin{itemize}
\item No funding was received for conducting this study.
\item The authors have no relevant financial or non-financial interests 
to disclose.
\item Ethics approval: Not applicable.
\item Consent to participate: Not applicable.
\item Consent for publication: Not applicable.
\item Availability of data and materials: Not applicable.
\item Code availability: Not applicable.
\item Authors' contributions: Both authors contributed equally to this work.
\end{itemize}

\ 

{\Large Note}

\ \ 

The expressions ``\textit{pure state" }and \textit{``mixture" (reiner Fall,
Gemenge) }may be found in (Weyl, 1927).\ The expressions \textit{reiner Fall}
and \textit{Gemenge} were not used in the 1932 book by von Neumann.

\ 

{\Large References}

\ \

Ballentine, Leslie E. 1998. \textit{Quantum Mechanics: A Modern Development}%
. Singapore: World Scientific.

Bell,\textit{\ }John Stewart.\ 1990. Against \textquotedblleft Measurement"%
\textit{. Proceedings of the NATO International school of History:
``Sixty-Two Years of Uncertainty", Erice, August 5-15,1989. }NATO ASI Series
B, \textit{Physics}, 226: 17-31, edited by\textit{\ }Miller A.I., Boston:
Springer.https://doi:10.1007/978-1-4684-8771-8\_3.

Bodor, Andr$\overset{,}{\text{a}}$s, and Lajos Di$\overset{,}{\text{o}}$si.
2012. Comment on ``Underlining some limitations of the statistical formalism
in quantum mechanics'" by Fratini and Hayrapetyan\textit{. }%
arXiv:1110.4549v2 [quant-ph].

Bricmont, Jean, and Sheldon Goldstein. 2018. Diagnosing the Trouble With
Quantum Mechanics\textit{. }arXiv: 1804.03401 [quant-ph].

Buchachenko, Anatoly L., F.I. Dalidchik, and B.R. Shub. 2001. Single spin
ESR. \textit{Chemical\ Physics Letters}. 340: 103-108\textit{.}

Deville, Yannick, and Alain Deville. 2007. Blind separation of quantum states: estimating two qubits from an isotropic Heisenberg spin coupling model, Proceedings of the 7th International Conference on Independent Component Analysis and Signal Separation (ICA 2007) (LNCS 4666, Springer-Verlag, London, 706 (2007)."

Deville, Yannick, and Alain Deville. 2012. Classical-processing and quantum-processing signal separation methods for qubit uncoupling, Quantum Inf. Process. 11, 1311-1347.
    
Deville, Yannick, and Alain Deville. 2014. Blind source separation: Advances in theory, algorithms and applications, Editors: Naik G.R., Wang W., Springer, Berlin, Germany, Ch. 1: Quantum-source independent component analysis and related statistical blind qubit uncoupling methods, 3-38 (2014).

Deville, Alain, and Yannick Deville. 2017. Concepts and Criteria for Blind Quantum Source Separation and Blind Quantum Process Tomography, \textit{Entropy}.

Deville, Yannick, and Alain Deville. 2015. From blind quantum source separation to blind quantum process tomography, in Proceedings of the 12th International Conference on Latent Variable Analysis and Signal Separation, LVA/ICA 2015 (Liberec, Czech Republic, Aug. 25-28, 2015), Springer International Publishing, Switzerland, LNCS 9237, 184--192 (Aug. 2015).

Deville, Yannick, and Alain Deville. 2017. The Blind Version of Quantum Process Tomography: Operating with Unknown Input Values, in Proceedings of the 20th World Congress of the International Federation of Automatic Control, (IFAC 2017, Toulouse, France, July 9-14, 2017).12228--12234 (2017).
    
Deville, Yannick, and Alain Deville. 2020. Quantum process tomography with unknown single-preparation input states: Concepts and application to the qubit pair with internal exchange coupling, Phys. Rev. A, 101, 042332.
   
Deville, Alain, and Yannick Deville. 2022. Random-coefficient pure
states and statistical mixtures. arXiv:2201.03248v2 [quant-ph].

Deville, Yannick, and Alain Deville. 2022. Beyond the density operator and $%
Tr(\rho A)$: Exploiting the higher-order statistics of random-coefficient
pure states for quantum information processing\textit{. }arXiv:2204.10031v1
[quant-ph].

Dirac, Paul A.M.\ 1939. A new notation for quantum mechanics\textit{.
Mathematical Proceedings of the Cambridge Philosophical Society}. 35:
416-418.

Fano, Ugo. 1957. Description of States in Quantum Mechanics by Density and
Operator Techniques\textit{. Reviews of Modern\ Physics}. 29: 74-93.

Fratini, Filippo, and Armen G. Hayrapetyan. 2011. Underlining some
limitations of the statistical formalism in quantum mechanics\textit{.} 
\textit{Physica Scripta.}\ 84: 035008 (5 pp).

Fratini, Filippo, and Armen G. Hayrapetyan. 2012. Underlining some
limitations of the statistical formalism in quantum mechanics: Reply to the
Comment of Bodor and Di$\overset{,}{\text{o}}$si. arXiv:1204.1071v1
[quant-ph].

Mathews, Jon, and R. L. Walker.1965. \textit{Mathematical Methods of Physics. }%
New York: Benjamin.

Nenashev, Alexei. 2016. Why state of quantum system is fully defined by
density matrix\textit{, }arXiv:1601.08205 [quant-ph].

Ohya, Masanori, and Igor Volovitch. 2011.\textit{\ Mathematical Foundations
of Quantum Information and Computation and Its Applications to Nano- and
Bio-systems. }Heidelberg: Springer.

Robertson, Howard Percy. 1929. The Uncertainty Principle, \textit{Physical
Review}, 34: 163-164.

Sakurai, Jun John. 1994. \textit{Modern Quantum Mechanics, }second edition,
Boston: Addison-Wesley.

Siomau, Michael, and Stephan U. Fritzsche. 2011. Quantum computing with mixed states, \textit{Eur. Phys. J. D}, 62, 449--456

Von Neumann, John. 1932. \textit{Mathematische Grundlagen der Quantenmechanik%
}. Berlin: Springer. English translation: 1955. \textit{Mathematical
Foundations of Quantum Mechanics: }Princeton: Princeton University Press.

Weinberg, Steven. \textit{\ }2017. The Trouble with Quantum Mechanics,%
\textit{The New York Review of Books.} January 19, 2017 Issue.

Weinberg, Steven. 2013. \textit{Lectures on Quantum Mechanics}. Cambridge:
Cambridge University Press.

Weyl, Hermann. 1927. Quantenmechanik und Gruppentheorie\textit{, Zeit. Physik%
} 46\textbf{: }1-46.

Zeh, Heinz-Dieter. 1970.\ On the Interpretation of Measurement in Quantum Theory. 
\textit{Foundations of Physics }1: 69-76.

\end{document}